\documentclass[twocolumn]{aastex631}
\usepackage{amsmath}
\usepackage{hyperref}
\usepackage{nameref}

\begin{document}

\title{A Short History of (Orbital) Decay: Roman's Prospects for Detecting Dying Planets}

\author[0009-0008-9182-7471]{Kylee Carden}
\email{carden.33@osu.edu}
\affiliation{Department of Astronomy, The Ohio State University, 140 West 18th Ave., Columbus, OH 43210 USA}

\author[0000-0003-0395-9869]{B. Scott Gaudi}
\affiliation{Department of Astronomy, The Ohio State University, 140 West 18th Ave., Columbus, OH 43210 USA}

\author[0000-0002-4235-6369]{Robert F. Wilson}
\affiliation{Department of Astronomy, University of Maryland, College Park, MD, USA}
\affiliation{NASA Goddard Space Flight Center, 8800 Greenbelt Rd, Greenbelt, MD, USA}



\begin{abstract}
The Roman Space Telescope Galactic Bulge Time Domain Survey (GBTDS) is expected to detect $\sim 10^5$ transiting planets. Many of these planets will have short orbital periods and are thus susceptible to tidal decay.  We use a catalog of simulated transiting planet detections to predict the yield of orbital decay detections in the Roman GBTDS.  Assuming a constant stellar tidal dissipation factor, $Q_{*}^{'}$, of $10^6$, we predict $\sim5-10$ detections.  We additionally consider an empirical period-dependent parameterization of $Q_{*}^{'}\propto P^{-3}$ and find a substantially suppressed yield.  We conclude that Roman will provide constraints on the rate of planet engulfment in the Galaxy and probe the physics of tidal dissipation in stars.
\end{abstract}



\section{Introduction} \label{sec:intro}
\par It has been known for some time that the excitation of tides by a satellite can result in orbital decay \citep{counselman_outcomes_1973}, which is thought to have played a role in the evolution of the Solar System \citep{goldreich_q_1966}. Until the discovery of the first exoplanet around a Sun-like star \citep{mayor_jupiter-mass_1995}, there was only the Solar System and its one formation outcome to study. Since the first observation of an exoplanetary transit \citep{charbonneau_detection_2000}, surveys have been conducted from the ground \citep[e.g.][]{Alonso_TrES1_2004,bakos_wide-field_2004, pollacco_wasp_2006,pepper_kilodegree_2007} and space \citep[e.g.][]{borucki_kepler_2010,howell_k2_2014,ricker_transiting_2015} to detect and characterize transiting planets. These surveys have detected thousands of short- and medium-period planets around hosts with a wide range of stellar properties. 
\par Several studies \citep[e.g.][]{maciejewski_planet-star_2018,patra_continuing_2020,adams_doomed_2024} have sought to detect exoplanet orbital decay, but there remain only two robust direct detections: WASP-12~b \citep{maciejewski_departure_2016,patra_apparently_2017} and Kepler-1658~b \citep{vissapragada_possible_2022}. Indirect evidence suggests that decay sculpts the population of exoplanets we observe. \cite{schlaufman_evidence_2013} argue hot Jupiters are found less frequently around subgiant stars where tidal effects have shorter timescales, though the number of hot Jupiters around red giant stars complicates this picture \citep{grunblatt_giant_2019}. Main sequence hot Jupiter host stars are younger than average, suggesting a hot Jupiter destruction mechanism \citep{hamer_hot_2019,miyazaki_evidence_2023}. A recent detection of a transient associated with a planetary engulfment \citep{de_infrared_2023} further motivates searches for orbital decay.
\par The Nancy Grace Roman Space Telescope is NASA's upcoming astrophysics flagship mission \citep{spergel_wide-field_2015}. Roman will conduct the Galactic Bulge Time Domain Survey (GBTDS)\footnote{\url{https://science.nasa.gov/mission/roman-space-telescope/galactic-bulge-time-domain-survey/}} as one of its Core Community Surveys to detect cold and free-floating planets using microlensing \citep{bennett_simulation_2002,penny_predictions_2019,johnson_predictions_2020} and complete a census of planets. The GBTDS will take advantage of Roman's wide field-of-view and efficiency to observe $\sim240\times10^6$ stars down to 25 mag in the broad (0.93 -- 2.00-micron) F146 filter. High-precision observations towards the Galactic bulge over a long temporal baseline will produce science beyond microlensing, including the detection of an estimated $\sim100,000$ warm and hot transiting planets \citep{bennett_simulation_2002,montet_measuring_2017,wilson_transiting_2023}. 
\par There is substantial historical precedence for the detection of transiting planets from microlensing surveys, including the first planet discovered by transits, OGLE-TR-56~b \citep{udalski_optical_2002}. While radial velocities were used to confirm this planet \citep{konacki_extrasolar_2003}, this will generally not be possible with Roman given the faintness of most of the observed stars. 
\par False positives are ubiquitous in transit surveys \citep{fressin_false_2013}, and robust vetting will be essential for Roman transiting exoplanet science. Various eclipsing binary (EB) configurations are especially nefarious astrophysical false positives \citep{cameron_extrasolar_2012}, as they can mimic transit signals. Techniques employed by the Kepler and TESS surveys \citep[e.g., centroid shifts, ellipsoidal variation, secondary eclipse detection, astrodensity profiling,][]{batalha_planetary_2013,sullivan_transiting_2015,kipping_characterizing_2014} could be used for vetting candidates. Chromaticity checks using the transit depth inferred from observations in Roman's secondary filter(s) could provide an additional means of validating short-period planet candidates.
\par Roman will expand transiting exoplanet science by 1) discovering planets in unexplored regions of the Galaxy, up to $\sim$20 kpc away, that span the thin disk, thick disk, and bulge and 2) expanding the known transiting planet population by at least an order of magnitude \citep{wilson_transiting_2023}. This will enable demographic studies that probe the entire Galactic population of planets. The detection of an unprecedentedly large sample of planets will enable searches for intrinsically rare phenomena, such as orbital decay.
\par Perhaps the largest uncertainty in estimating a precise orbital decay yield comes from the stellar tidal dissipation factor $Q_{*}^{'}$, whose estimates can vary by orders of magnitude \citep[e.g.][]{ogilvie_tidal_2007} from approximately $10^{5.5}$ to $>\!10^8$ \citep{adams_doomed_2024}. The value of $Q_{*}^{'}$ depends on the physical processes by which energy dissipates \citep{2008EAS....29...67Z}. Previous work has suggested $Q_{*}^{'}$ depends on stellar mass, with the decay time only being short on the subgiant branch for the more massive stars \citep{weinberg_orbital_2024}, and on the tidal forcing frequency \citep[i.e. the orbital period, see, e.g.,][]{ogilvie_tidal_2007,collier_cameron_hierarchical_2018}. 
\par In this work, we consider the prospects for detecting orbital decay with the GBTDS. In Sec. \ref{sec:roman}, we lay out the features of the Roman Space Telescope, the Galactic Bulge Time Domain Survey, and a catalog of simulated transiting planet detections. In Sec. \ref{sec:meth}, we introduce formalism to estimate the signal of decay $dP/dt$ in each simulated system and the accuracy with which Roman can measure it. In Sec. \ref{sec:res}, we use our formalism to address survey optimization and extensions and to estimate the yield of Roman orbital decay detections.

\section{Roman and the GBTDS} \label{sec:roman}
\par The GBTDS will be carried out by the Wide Field Instrument (WFI) \citep{spergel_wide-field_2015}. The technical specifications of the WFI are described in \cite{akeson_wide_2019}, and the most up-to-date values are listed on a NASA webpage.\footnote{\url{https://roman.gsfc.nasa.gov/science/WFI_technical.html}} The WFI will leverage Roman's 2.4-m aperture primary mirror and 18 H4RG-10 4K x 4K detectors with broad near-infrared (NIR) sensitivity to achieve a field of view of 0.281 deg$^2$, $\sim200$ times larger than Hubble's WFC3, while maintaining a similar resolution of $0.11''$/pixel \citep{mosby_properties_2020}. The WFI instrument includes eight overlapping filters spanning 0.48 -- 2.3 micron. The primary F146 survey filter spans 0.93 -- 2.00 micron with a PSF FWHM of 0.105 arcsec.
\par The Roman Observations Time Allocation Committee provided their recommendations for the strategies for the Core Community Surveys\footnote{\url{https://roman.gsfc.nasa.gov/science/ccs/ROTAC-Report-20250424-v1.pdf}}. These recommendations differ from the notional survey defined in \cite{penny_predictions_2019} in numerous ways, though most of these changes are of minimal consequence for orbital decay science. We proceed considering the notional survey, which has the following specifications:
\begin{enumerate}
   \item six 72-day observing seasons centered on either the autumnal or vernal equinoxes when the Galactic bulge is visible from L2;
   \item a survey baseline of 5 years with the observing seasons arranged as [1, 1, 1, 0, 0, 0, 0, 1, 1, 1], 1 = ON, 0 = OFF;
   \item a 15-minute observing cadence (47 s exposures) for the primary filter F146 during each observing season; and
   \item seven adjacent observing fields situated a couple of degrees from the Galactic Center
\end{enumerate}
\par The Survey Definition Committee has laid out three distinct observing scenarios (underguide, nominal, and overguide). The number and placement of fields, season length, and cadence varies for each scenario. Since recent work advocated for changes to the season configuration \citep[e.g.][]{gould_one_2024} and cadence \citep[e.g.][]{kupfer_continuous_2023}, we focus on the ramifications of these two survey parameters for detecting orbital decay in Sec. \ref{sec:surveyopt}. We additionally consider the usefulness of extended observations for confirming decay candidates identified in the GBTDS.
\par \cite{wilson_transiting_2023} performed a pixel-level simulation of a single field of the \cite{penny_predictions_2019} notional survey to estimate a GBTDS yield of $\sim60,000$ -- $200,000$ transiting planets. This simulation includes $\sim 59\times10^{6}$ stars with F146 $<$ 21 as possible transiting planet hosts and led to a catalog of $\sim9,000$ detected planets in one field. \cite{wilson_transiting_2023} followed the approach of \cite{barclay_revised_2018} to simulate the exoplanet population, drawing the number of planets per star from a Poisson distribution consistent with estimated occurrence rates from \cite{hsu_occurrence_2019} and then assigning each planet its properties and a host star. Only main-sequence and subgiant stars with $\log{g}\ge 3.5$ were considered as viable planet hosts.
\par The catalog we use did not scale planet occurrence rates with metallicity, whose relationship is well-studied (e.g., \citealt{Gonzalez:1997,Santos:2001,Fischer:2005,Buchhave:2014}), albeit not well characterized for all planet populations. When accounting for the higher metallicities of the inner Galaxy, \cite{wilson_transiting_2023} found overall yield increases by a factor of $\sim3$, the yield of $4-8\,R_\oplus$ planets increases by $\sim5$, and the yield of $>8\,R_\oplus$ planets increases by $\sim2$. The detections of orbital decay we find are of giant planets, and thus if the scaling of planet frequency with metallicity adopted by \citet{wilson_transiting_2023} holds for the stellar populations probed by Roman, we can expect a yield of decaying planets that is a few times larger than our baseline predictions.

\section{Methods} \label{sec:meth}
\subsection{Transit Time Modeling}
\par In order to detect orbital decay, a quadratic ephemeris model with a shrinking period must be statistically preferred over the simpler linear ephemeris model. An approximate expression for period decay \citep{goldreich_q_1966,adams_doomed_2024} is
\begin{equation}
\label{eq:decay}
    \frac{dP}{dt} = -\left(\frac{27\pi}{2}\right)\left(\frac{M_p}{M_*}\right)\left(\frac{a}{R_*}\right)^{-5}\left(Q_{*}^{'}\right)^{-1},
\end{equation}
where $M_p$ is the planet mass, $M_*$ is the stellar mass, $a$ is the orbital semi-major axis, $R_*$ is the stellar radius, and $Q_{*}^{'}$ is the stellar tidal dissipation factor.
The linear ephemeris model is
\begin{equation}
    \label{eq:lin}
    T(E) = T_{0,lin} + P_{lin}\times{E},
\end{equation}
where $T(E)$ is the transit midpoint time predicted at epoch $E$, $T_{0,lin}$ is the reference ephemeris for $E=0$, and $P$ is the orbital period. The quadratic ephemeris model is
\begin{equation}
    \label{eq:quad}
    T(E) = T_{0,quad} + P_{quad}\times{E} + \frac{1}{2}\frac{dP}{dE}\times{E^2},
\end{equation}
where $T_{0,quad}$ is the reference ephemeris, and $dP/dE$ is the derivative of the period $P_{quad}$.
\par With the catalog of simulated detected planets from \cite{wilson_transiting_2023} and an assumption about the stellar tidal dissipation factor $Q_{*}^{'}$, we can estimate the rate of orbital decay experienced by each system. We then need to estimate how well Roman will be able to constrain $dP/dt$.
\subsection{Transit Time Covariance Matrix}
\par To estimate the uncertainty with which we can constrain $dP/dt$, we follow the formalism laid out in \citet{gould_chi2_2003} and \citet{gould_one_2024}. To make the analytic derivation simpler, we imagine observing one transit epoch at the central time of each season. Since the length of seasons (72 days) is small compared to the survey baseline (4.7 years), this simplification introduces only a small error, as we discuss in Sec. \ref{sec:valid}. We discuss an alternative season configuration in Sec. \ref{sec:surveyopt}. Considering each season as a single observation translates to observing epochs
\begin{equation}
    \label{eq:epoch}
    E_k
    =
    \begin{bmatrix}
    -9, & -7, & -5, & +5, & +7, & +9
    \end{bmatrix}
    \frac{L}{P},
\end{equation}
where $L\equiv\textnormal{yr}/{4}$ is half the minimum length of time between two observing season centers, $P$ is the orbital period, and the epoch $E=0$ is placed at the middle of the survey baseline.
\par We fit the observed transit times with a polynomial
\begin{equation}
    F(E; a_i,...,a_r) = \sum_{i=0}^{r} a_if_i(E),
\end{equation}
where the trial functions $f_i(E),...,f_r(E)$ are terms from the truncated Taylor series, $f_{i}(E) = E^{i}/i!$,
where $r=1$ for the linear fit and $r=2$ for the quadratic fit. In the quadratic case, we are fitting for coefficients $a_i$, where $a_0 = T_{0,quad}$, $a_1 = P_{quad}$, and $a_2 = dP/dE$ from Eq. \ref{eq:quad}.
The entries of the inverse covariance matrix of the fit are then
\begin{equation}
\label{eq:fisherentries}
    b_{ij} = \sum_{k=1}^{N} \frac{1}{\sigma_0^2}\frac{\partial F}{\partial a_i}\frac{\partial F}{\partial a_j}
    =\sum_{k=1}^{N} \frac{f_{i}(E_{k})f_{j}(E_{k})}{\sigma_{0}^2}
\end{equation}
with uncorrelated and constant uncertainties $\sigma_0$ in transit time measurements for each of the $N=6$ seasons of observing.
\par Because the defined epochs in Eq. \ref{eq:epoch} are distributed symmetrically about $E=0$, any matrix element $b_{ij}$ where $i+j$ is odd is exactly $0$. This leads to an inverse covariance matrix of
\begin{equation}
    b \equiv c^{-1} = \frac{N}{\sigma_{0}^2}
    \begin{bmatrix}
    1 & 0 & \langle E^2 \rangle/2 \\
    0 & \langle E^2 \rangle & 0 \\
    \langle E^2 \rangle/2 & 0 & \langle E^4 \rangle/4 \\
    \end{bmatrix},    
\end{equation}
where 
\begin{equation}
    \label{eq:disp}
    \langle E^t \rangle \equiv \frac{1}{N} \sum_{k} E_{k}^{t}.  
\end{equation}
This can be inverted to obtain the covariance matrix of
\begin{equation}
    \label{eq:covar}
    c = \frac{\sigma_{0}^2}{N}
    \begin{bmatrix}
    \frac{\langle E^4 \rangle}{\langle E^4 \rangle - \langle E^2 \rangle^{2}} & 0 & \frac{2\langle E^2 \rangle}{\langle E^2 \rangle^2-\langle E^4 \rangle} \\
    0 & \frac{1}{\langle E^2 \rangle} & 0 \\
    \frac{2\langle E^2 \rangle}{\langle E^2 \rangle^2-\langle E^4 \rangle} & 0 & \frac{4}{\langle E^4 \rangle - \langle E^2 \rangle^{2}} \\
    \end{bmatrix}.    
\end{equation}
\subsection{Analytic Uncertainty}
Following the definition in Eq. \ref{eq:disp}, we find that the epoch averages are
\begin{equation}
    \label{eq:values}
    \left\langle \left(E\frac{P}{L}\right)^{2} \right\rangle = \frac{155}{3}\;\textnormal{and}\;  \left\langle \left(E\frac{P}{L}\right)^{4} \right\rangle = \frac{9587}{3}.
\end{equation}
The uncertainty in $T_0$ is 
\begin{equation}
    \sigma_{T_0} = \sqrt{c_{00}} = \frac{\sqrt{9587}}{16\sqrt{37}}\sigma_{0} \simeq 1.01\sigma_0.
\end{equation} 
The uncertainty in $P$ is
\begin{equation}
    \sigma_{P} = \sqrt{c_{11}} = \frac{1}{\sqrt{310}}\frac{P}{L}\sigma_{0} \simeq 0.057\frac{P}{L}\sigma_{0} .
\end{equation}
The uncertainty in $dP/dE$ is
\begin{equation}
    \label{eq:epochunc}
    \sigma_{dP/dE} = \sqrt{c_{22}} = \frac{\sqrt{3}}{8\sqrt{37}}\frac{P^2}{L^2}\sigma_0 \simeq 0.036 \frac{P^2}{L^2}\sigma_{0}.
\end{equation}
\par The period derivative $dP/dt = P^{-1} dP/dE$ depends on both $P$ and $dP/dE$, and the uncertainty in $dP/dt$ has contributions from both terms. However, the ratio of relative uncertainties is
\begin{equation}
    \left(\frac{\sigma_{dP/dE}}{dP/dE}\right) \times \left(\frac{\sigma_P}{P}\right)^{-1} \simeq 0.6 \frac{P}{LdP/dt}.
\end{equation}
Since the orbital period changes we are searching for are very small (i.e. $L dP/dt \ll P$), only the uncertainty in $dP/dE$ contributes appreciably to uncertainty in $dP/dt$. This leads us to the expression
\begin{equation}
    \sigma_{dP/dt} \approx \frac{\sqrt{3}}{8\sqrt{37}}\frac{P}{L^2}\sigma_{0}.
\end{equation}
\par We adopt the analytic transit time uncertainty of
\begin{equation}
    \sigma_{t_c} = Q^{-1}T\sqrt{\theta/2}    
\end{equation}
from \cite{carter_analytic_2008}, where
\begin{equation}
    Q\equiv\frac{\delta}{\sigma_{phot}}\sqrt{\Gamma{T}}
\end{equation}
is the total signal-to-noise ratio of the transit in the limit $r\to0$, $\theta\equiv\tau/T$ is the ratio of ingress/egress time to the total transit time $T$, $\Gamma$ is the sampling rate or inverse cadence, and $\sigma_{phot}$ is photometric uncertainty per exposure. Roman's $\sigma_{phot}$ is shown in Fig. 4 of \cite{wilson_transiting_2023}, and for stars with an F146 mag $\sim18$, $\sigma_{phot}\approx 0.2\%$  To first order, $\theta\sim\sqrt{\delta}$ so
\begin{equation} \label{eq:sigmatcapprox}
\sigma_{t_c}\approx\sigma_{phot}\sqrt{\frac{T}{2\delta^{3/2}\Gamma}},
\end{equation}
The transit timing uncertainty for a season from Eq. \ref{eq:fisherentries} is $\sigma_0 = \sigma_{t_c}/\sqrt{n_s}$, where $n_s$ is the number of transits observed per season, leading to an uncertainty in the period derivative of
\begin{equation}
\label{eq:finalsigma}
    \sigma_{dP/dt} \approx \frac{3}{8\sqrt{37}}\frac{P}{L^2}\sqrt{\frac{T}{\Gamma\delta^{3/2}}}\frac{\sigma_{phot}}{\sqrt{n_{tot}}},
\end{equation}
where $n_{tot} = 6n_s$ is the total number of transits observed over the duration of the survey.

\subsection{Validation}
\label{sec:valid}
\par Assuming that we observe only six transits at the middle of each season does introduce a small error in estimating the epoch averages. We are making the approximation that
\begin{equation}
    \left\langle \left(E\frac{P}{L}\right)^{t} \right\rangle \simeq \left(\frac{P}{L}\right)^{t}\frac{1}{M}\sum_{i=1}^{M} E_{i}^{t},
\end{equation}
where $M$ is the full number of epochs that would actually be observed by Roman. Both of these approximations are good to less than $1\%$ for periods $<5$ days because the observing seasons are short compared to the survey baseline. This suggests we can estimate the uncertainty for each parameter accurately.
\begin{figure}[!t]
\centering
\includegraphics[width=0.47\textwidth]{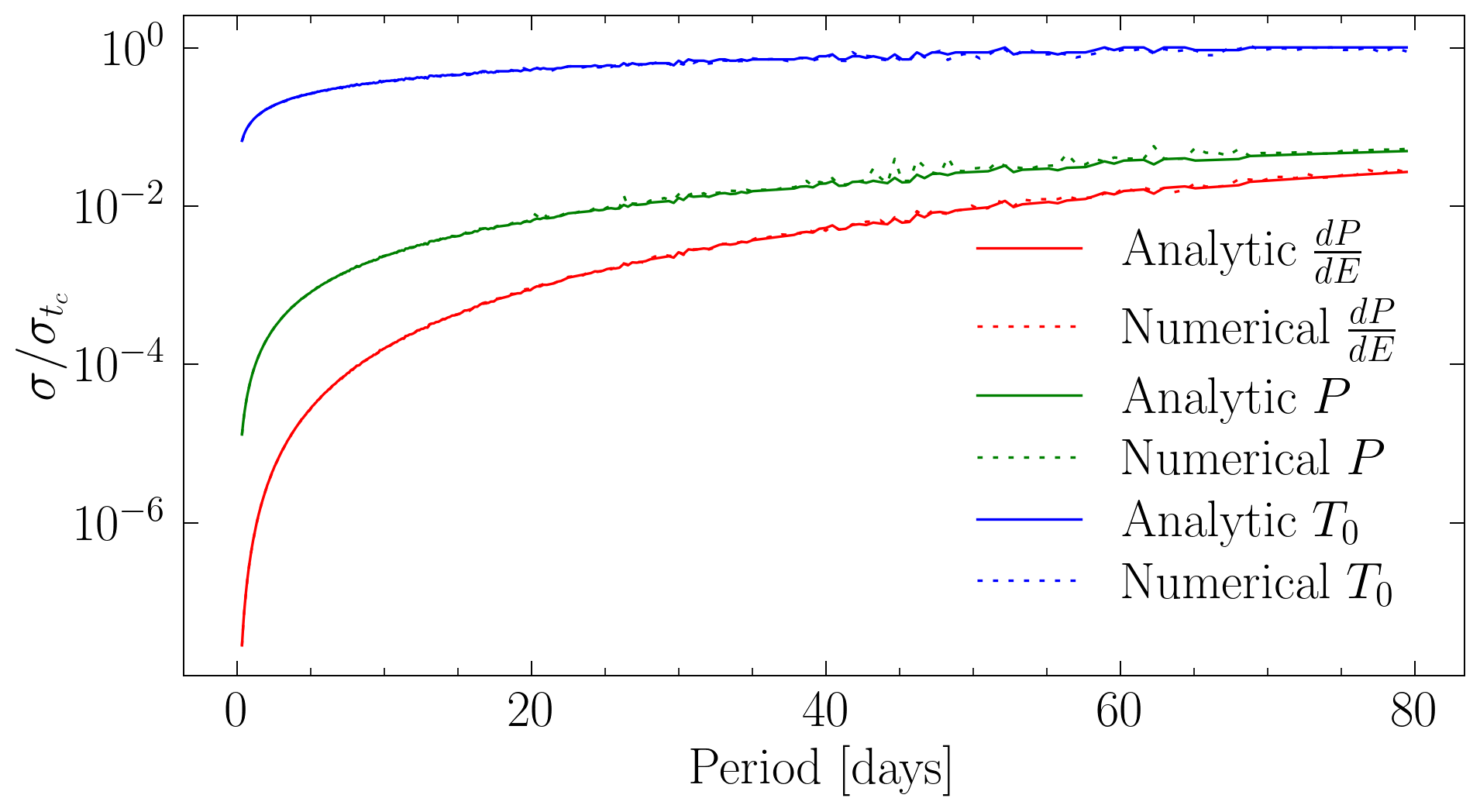}
\caption{Comparison of the analytic estimate of uncertainty with the numerical results as a function of period, showing good agreement for all three parameters in the short-period regime.}
\label{fig:sigma}
\end{figure}
\par To check our analytic result, we simulate transits of planets of various periods with an injected decay signal, i.e., a period derivative. We keep only the transits that would be observed in GBTDS observing seasons. Then, we perform a quadratic fit, evaluate the uncertainty in $T_0$, $P$, and $dP/dE$, all relative to $\sigma_{t_c}$, and find good agreement between the analytic estimates and the numerical results, as shown in Fig. \ref{fig:sigma}.

\section{Estimating the Yield} \label{sec:res}
\subsection{Survey Optimization and Extensions} 
\label{sec:surveyopt}
\par Although the GBTDS survey strategy is essentially defined, we investigate the implications of survey parameters on orbital decay science. \cite{gould_one_2024} suggested an alternative season configuration ([1, 1, 0, 0, 1, 1, 0, 0, 0, 1, 1], 1 = ON, 2 = OFF) better optimized for black hole science. This configuration translates to observing epochs
\begin{equation}
    \label{eq:epochg}
    E_{k,\,G}
    =
    \begin{bmatrix}
    -9, & -7, & -1, & +1, & +7, & +9
    \end{bmatrix}
    \frac{L}{P}.
\end{equation}
Following the definition in Eq. \ref{eq:disp}, we find that
\begin{equation}
    \label{eq:valuesg}
    \left\langle \left(E\frac{P}{L}\right)^{2} \right\rangle_G = \frac{131}{3}\;\textnormal{and}\;  \left\langle \left(E\frac{P}{L}\right)^{4} \right\rangle_G = \frac{8963}{3}.
\end{equation}
Recalling Eq. \ref{eq:epochunc}, we can compare the uncertainty with which we can measure $dP/dE$ 
\begin{equation}
    \frac{\sigma_{dP/dE}^G}{\sigma_{dP/dE}} = \sqrt{\frac{\langle{E^4}\rangle - \langle{E^2}\rangle^2}{\langle{E^4}\rangle_{G} - \langle{E^2}\rangle_{G}^2}} = \sqrt{\frac{37}{76}} \simeq 0.70,
\end{equation}
using the values from Eq. \ref{eq:valuesg}. The Gould configuration leads to an uncertainty in $dP/dt$ that is 70\% of the uncertainty in the notional survey. \cite{gould_one_2024} similarly found this improvement for measurements of astrometric accelerations with Roman.
\par The effects of other survey parameters are more subtle since they might also appreciably affect the number of transiting planets detected in addition to the uncertainty with which we can measure orbital decay. The cadences under consideration range from 12.1 to 14.8 min, suggesting a modest $\sim10\%$ effect on measuring uncertainty in $dP/dt$ due to differences in the cadence at fixed exposure time.  However, the decreased cadence results in an increase in the exposure time at fixed overhead time, leading to a further improvement in the uncertainty in $dP/dT$. The Survey Definition Committee ultimately recommended $5-7$ primary bulge fields plus a field at the Galactic Center.  For large radii, high signal-to-noise ratio detections, this yield is approximately proportional to the number of sources, and thus going from 7 to 5 fields would decrease the yield by $\sim30\%$. Assuming nothing about the properties of each field, each additional field simply increases the probability of finding a high-significance decay system because there are more systems in total.
\par \cite{2024arXiv240614531G} advocated for doubling the sampling rate relative to the baseline Penny et al. (2019) survey design to $\Gamma = 8/{\rm hr}$ (or one observation every 7.5 minutes), in order to increase the yield of low-mass free-floating planets detected by microlensing.  This would require halving the number of fields, and would result in an  increase in the yield of planets with small radii, but a decrease in the yield of planets with larger radii.  If there exists a large population of small, short-period planets that are just below detectability with the nominal cadences of every 12.4 to 14.8 minutes, then this strategy could increase the yield of tidal decay detections dramatically.  However, the Survey Definition Committee ultimately did not consider such high cadences.  This may be an interesting strategy to consider in future campaigns during an extended Roman mission.
\par For survey extension to help, we must ensure the ephemerides remain accurate since we must know the epoch $E$ corresponding to any transit. The number of orbits before the ephemeris is degraded is approximately
\begin{equation}
    \frac{P}{\sigma_P}\approx\sqrt{\frac{155}{3}}L\frac{\sqrt{n_{tot}}}{\sigma_{t_c}} 
\end{equation}
For short-period planets orbiting bright stars, $n_{tot}>100$ and $\sigma_{t_c}<100$ min. This means that observations taken even years after the completion of the primary survey can be used to follow up decay candidates.
\par
\begin{figure}[!t]
\centering
\includegraphics[width=0.47\textwidth]{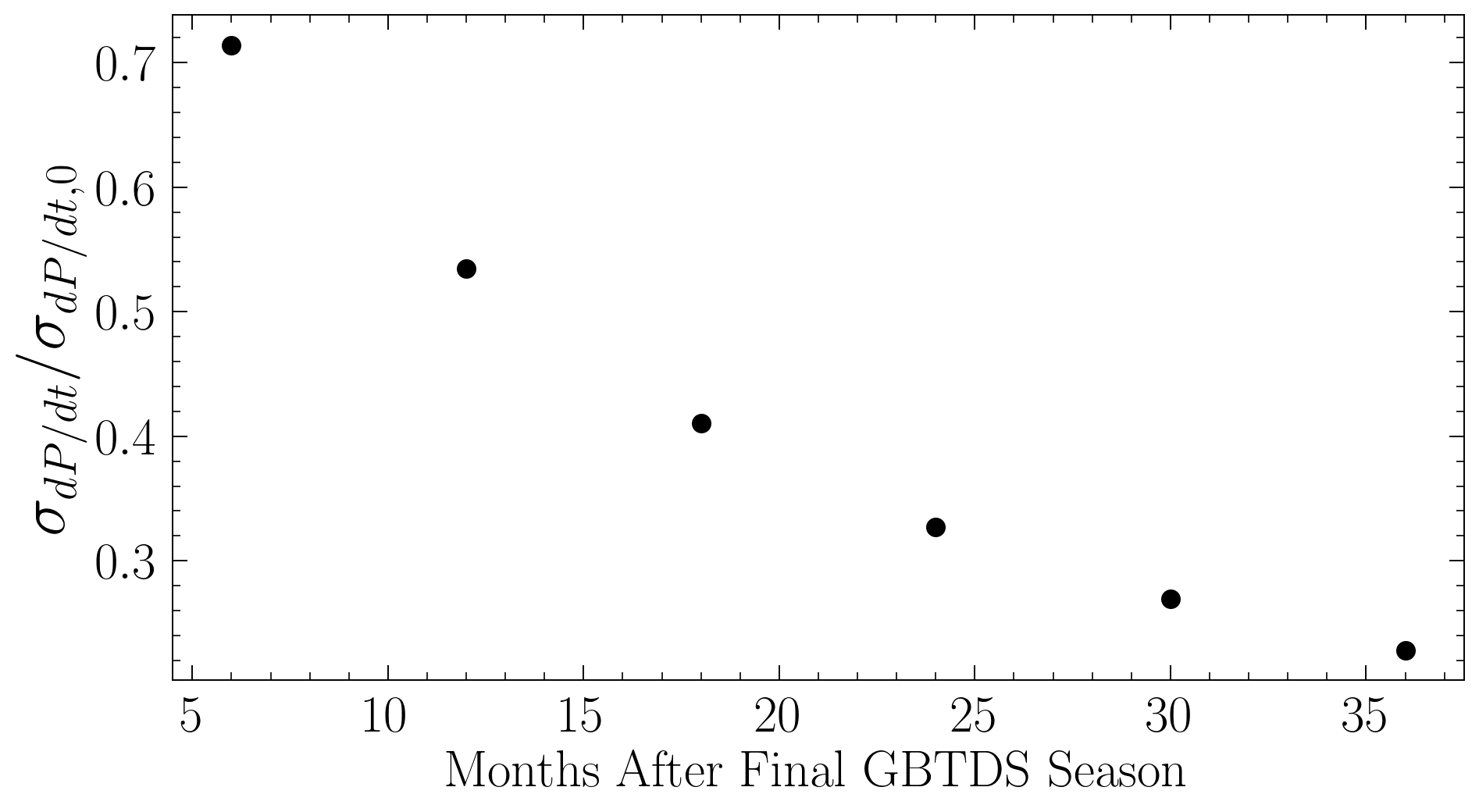}
\caption{The reduction in uncertainty, relative to the GBTDS uncertainty $\sigma_{dP/dt,0}$, that can be achieved with an additional season of observing carried out some number of months after the conclusion of the GBTDS.}
\label{fig:extend}
\end{figure}

 Since adding an additional season breaks the symmetry we used to construct the covariance matrix in Eq. \ref{eq:covar}, we numerically assessed the impact of adding an additional season. Season midpoints must be separated by at least six months given the bulge visibility windows. Fig. \ref{fig:extend} shows the reduction in $\sigma_{dP/dt}$ that can be achieved with a single additional season whose midpoint is placed in a multiple of 6 months after the conclusion of the GBTDS. Additional observations after a few years have passed significantly improve our ability to measure $dP/dt$.

\subsection{Catalog Detections}
\par
\begin{figure}[!b]
\centering
\includegraphics[width=0.47\textwidth]{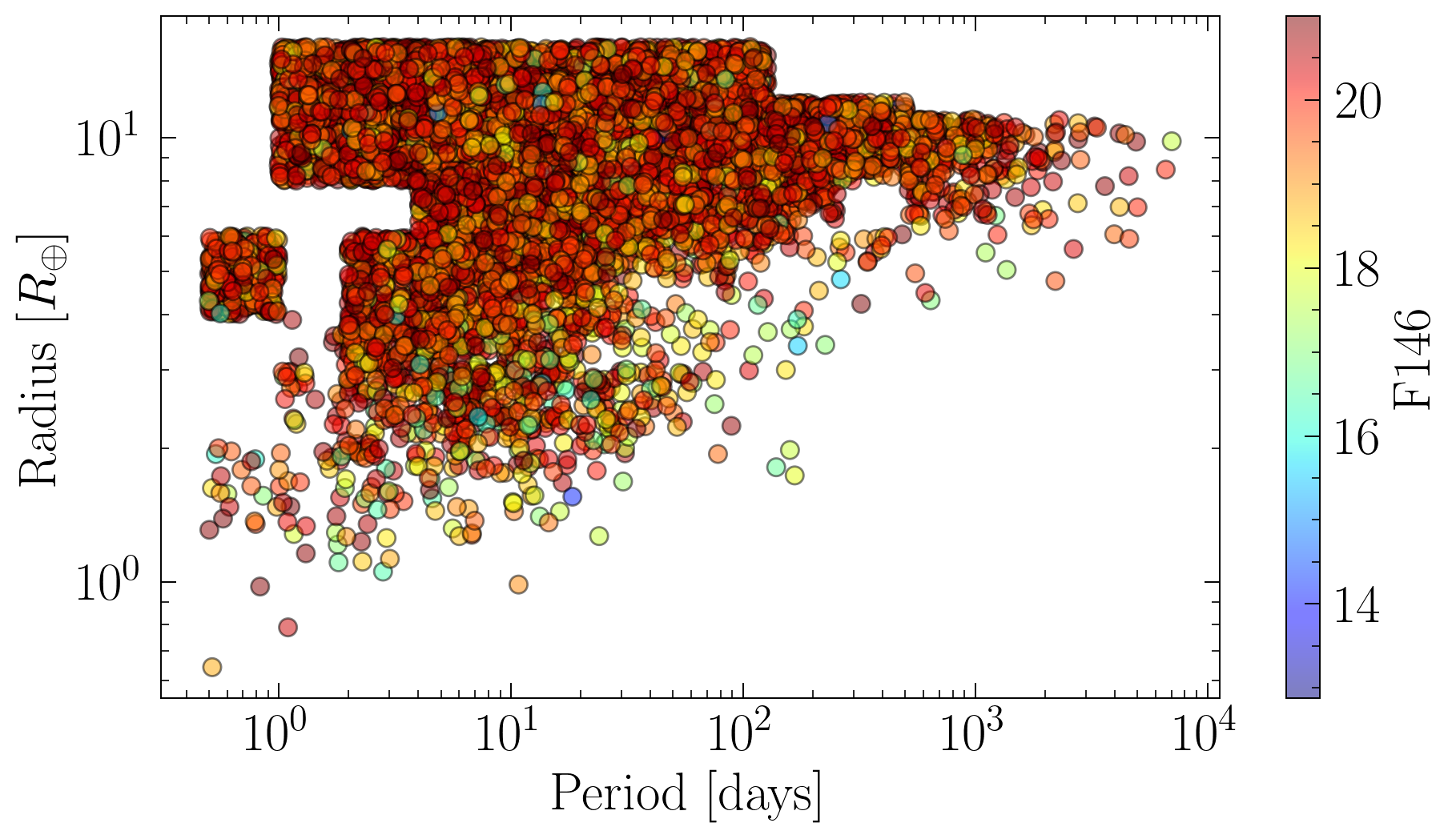}
\caption{The distribution of the planets from the \cite{wilson_transiting_2023} simulated planet catalog in period and radius. The color bar indicates the apparent magnitude of each planet's host star.}
\label{fig:catalog}
\end{figure}

 We analyzed the catalog of simulated planet detections from \cite{wilson_transiting_2023}. It provides a planetary mass using the mass-radius relationship of \citet{chen_probabilistic_2017}, stellar mass, planetary orbital period, stellar radius, planetary radius, transit duration, diluted transit depth, F146 magnitude, and number of transits observed for each simulated detection. It simulated a single field of transiting planet detections in the GBTDS. Fig. \ref{fig:catalog} shows the 9279 simulated detections in the space of period and radius. We find a few dozen planets within the Roche radii of their host stars and eliminate them from further analysis. We additionally assume eccentricity $e=0$ in all cases.
We use a noise model produced by The Roman IMage and TIMe-series SIMulator\footnote{\url{https://github.com/robertfwilson/rimtimsim}}.
\begin{figure}[!t]
\centering
\includegraphics[width=0.47\textwidth]{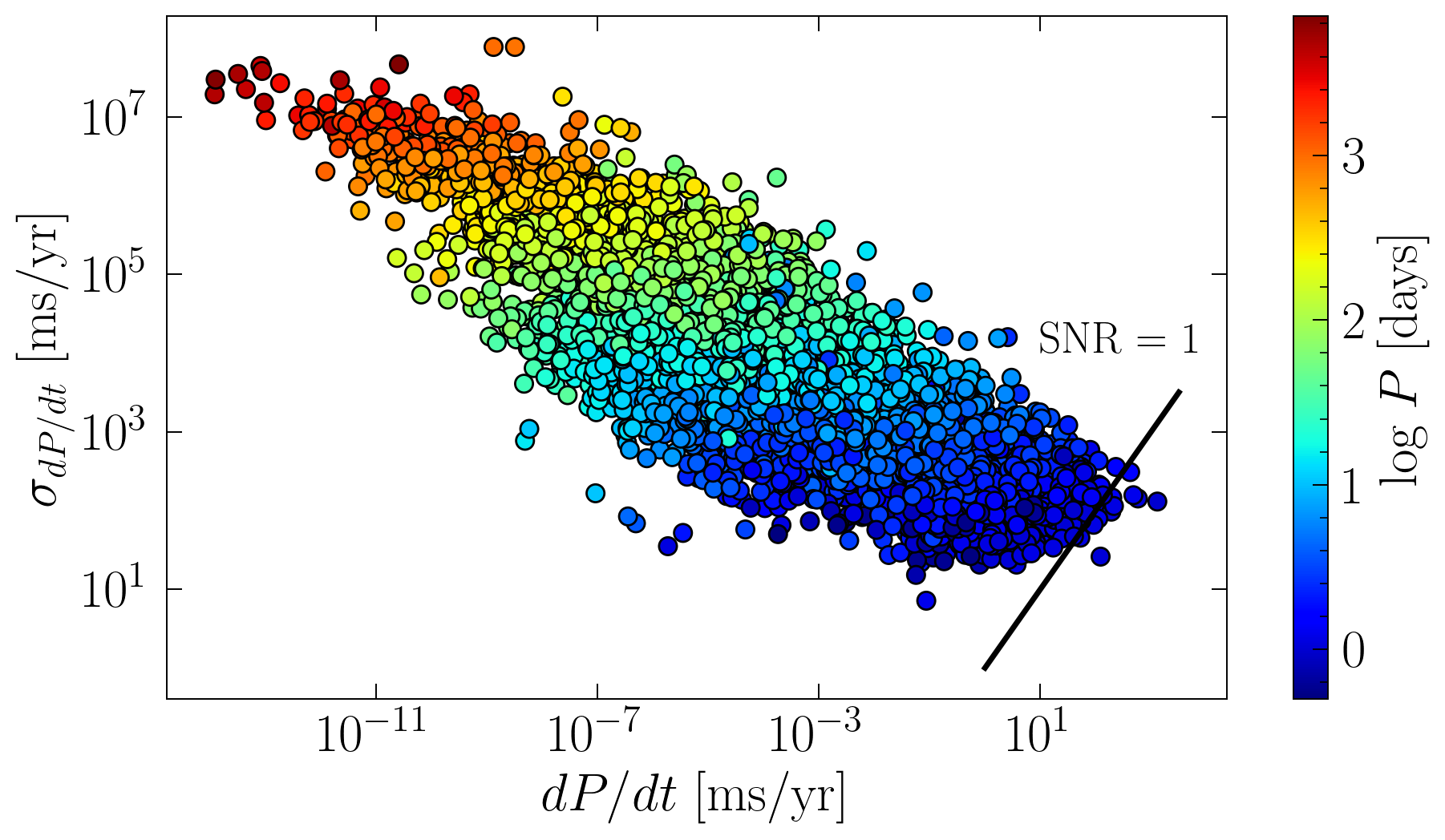}
\caption{The distributions of $dP/dt$ and $\sigma_{dP/dt}$ for each simulated planet detection. These quantities are highly correlated due to the strong dependence of each on the planet's orbital period, as shown by the color bar. WASP-12~b has ${dP/dt}\sim30$ ms/yr.}
\label{fig:dist}
\end{figure}

\par To estimate the yield of orbital decay detections, we need a model for the stellar tidal dissipation factor $Q_{*}^{'}$, since  there is little consensus on the mechanism or efficiency of tidal dissipation. There are several parametrizations of $Q_{*}^{'}$ that scale with the planet's orbital period and occasionally other system properties \citep[e.g][]{essick_orbital_2016,penev_empirical_2018,millholland_empirical_2025}. For our baseline estimate, we assumed a constant $Q_{*}^{'}=10^{6}$. Fig. \ref{fig:dist} shows the resulting distributions of period derivatives $dP/dt$ using Eq. \ref{eq:decay}  and their uncertainties using Eq. \ref{eq:finalsigma}.

\begin{deluxetable}{r|cccc}[!b]
\tablecolumns{4}
\tablewidth{1.0\columnwidth} 
\tablecaption{High-Significance Systems} 
\tablehead{Quantity & \#1923 & \#1399 & \#6041 & \#6286}
\startdata
    $M_{*}$ [$M_{\odot}$] & 1.1 & 1.5 & 1.1 & 1.1 \\
    $R_*$ [$R_{\odot}$] & 2.7 & 2.3 & 3.0 & 2.5 \\
    F146 & 17.8 & 15.9 & 17.9 & 17.7 \\ 
    $M_p$ [$M_\oplus$] & 3930 & 1780 & 1430 & 4500 \\
    $R_p$ [$R_{\oplus}$] & 11.7 & 13.0 & 13.5 & 10.8 \\
    $P$ [days] & 1.00 & 1.03 & 1.07 & 1.23 \\
    $dP/dt$ [ms/yr] & 1290 & 122 & 576 & 478 \\
    $(dP/dt)/\sigma$ & 10 & 4.7 & 4.1 & 3.1 \\
\enddata  
\label{tab:systems}
\tablecomments{Each simulated planet is identified with its index in the catalog.}
\end{deluxetable}
\par We then calculated the orbital decay signal-to-noise ratio, SNR $= (dP/dt)/(\sigma_{dP/dt})$ for each planet in the catalog. The system properties of the four highest SNR examples are reported in Table \ref{tab:systems}. They are massive planets orbiting bright stars with period derivatives generally much larger than those inferred for WASP-12~b and Kepler-1658~b.
\begin{figure}[!t]
\centering
\includegraphics[width=0.47\textwidth]{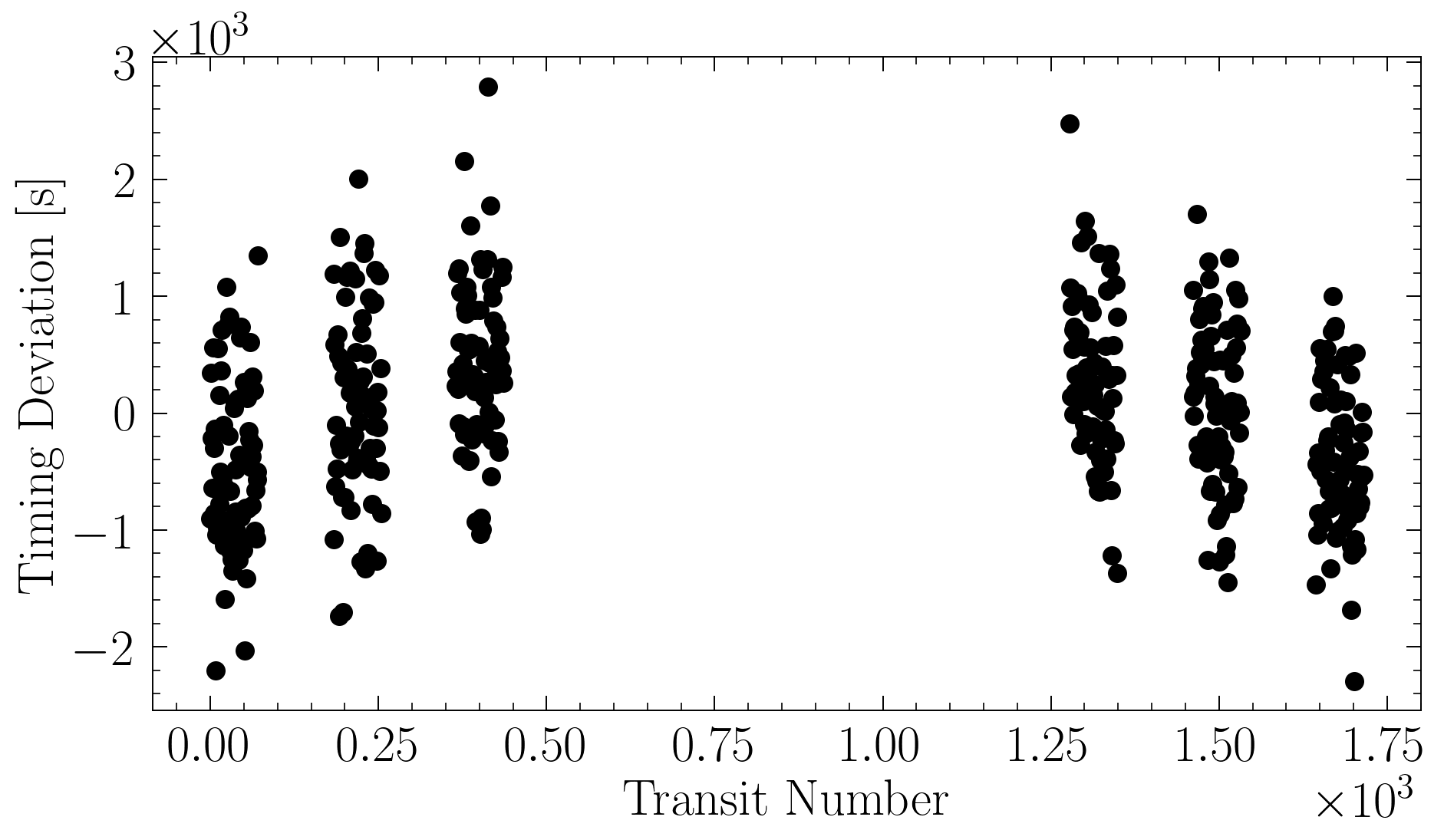}
\caption{Example transit timing for planet \#1923 in Table \ref{tab:systems} from the GBTDS using the \cite{penny_predictions_2019} notional survey. This planet has the most robust detection of orbital decay in the simulated catalog.}
\label{fig:decayex}
\end{figure}
\par To show what transit timing signal would look like with Roman, we simulated the detections of planet \#1923 from Table \ref{tab:systems} including the effect of decay from the $dP/dt$ we estimated. We added Gaussian noise for the transit times and subtracted the best-fit linear model. The quadratic residuals that signal the decay of the orbit are shown in Fig. \ref{fig:decayex}.
\begin{figure}[!b]
\centering
\includegraphics[width=0.47\textwidth]{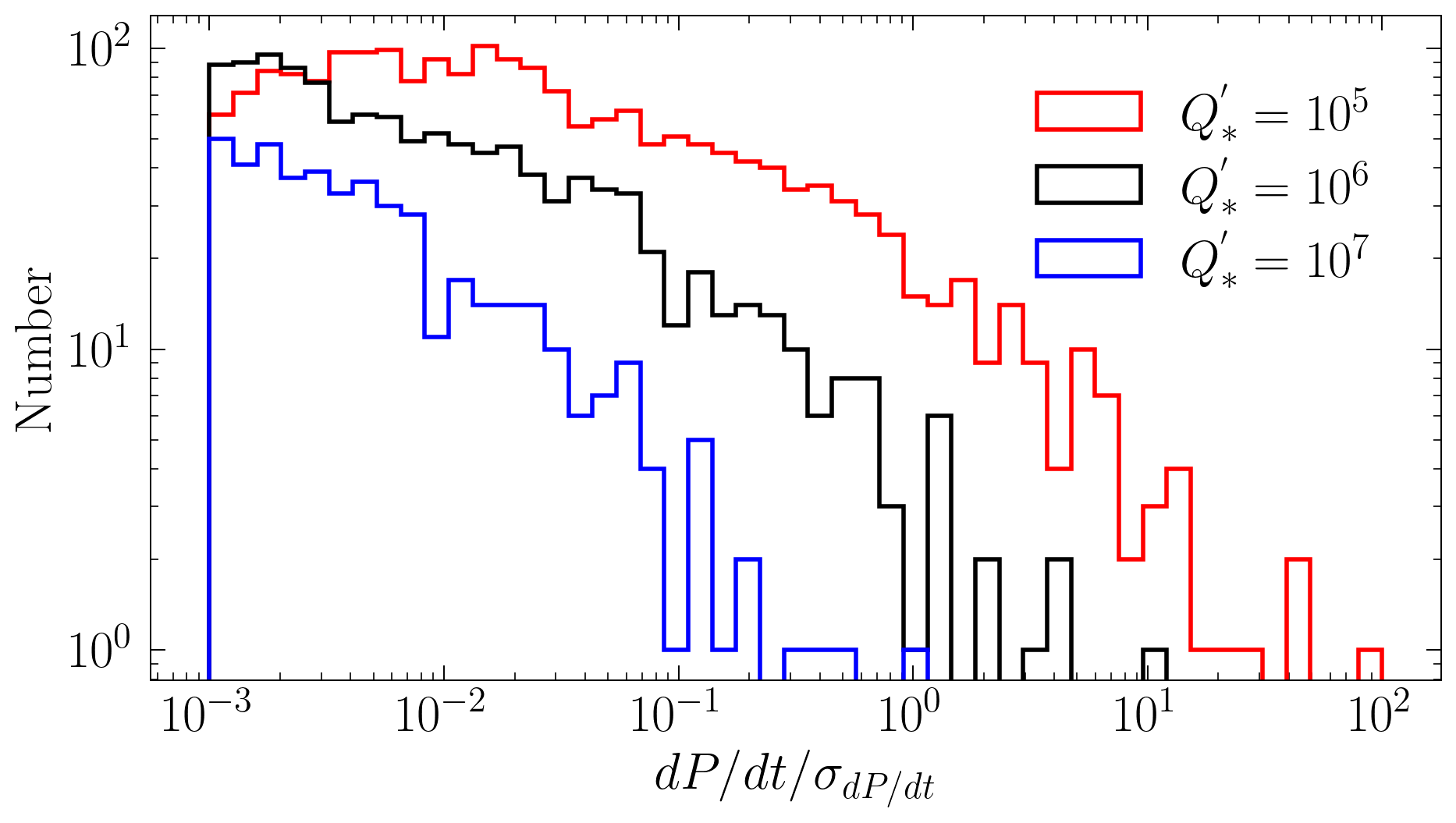}
\caption{The histogram of orbital decay SNRs $>10^{-3}$ using three distinct values of $Q_{*}^{'}$.}
\label{fig:hist}
\end{figure}
\par Fig. \ref{fig:hist} shows the distribution of the orbital decay signal-to-noise ratios $>10^{-3}$ for planets in the catalog using our baseline estimate assuming $Q_{*}^{'}=10^6$ and alternative scenarios $Q_{*}^{'}=10^5$ and $Q_{*}^{'}=10^7$. Since Roman will find $\sim100,000$ transiting planets, we must set a SNR threshold such that it is unlikely random fluctuations produce statistical false positives. SNR $= (dP/dt)/\sigma \simeq \sqrt{\Delta\chi^2},$ where $\Delta\chi^2$ compares the goodness-of-fit of the linear and quadratic models. $\Delta\chi^2$ follows a $\chi^2$ distribution with one degree of freedom assuming uncorrelated, Gaussian noise. $P(\Delta\chi^2>19.5)\simeq 10^{-5},$ suggesting SNR $\simeq\sqrt{19.5}\simeq4.4$ is a reasonable threshold for detection. We thus expect that $1-2$ instances of orbital decay would be detectable for this catalog assuming $Q_{*}^{'}=10^6$.  
\par Two detections out of the $\sim3600$ hot Jupiters ($P<10\,$days and $R_p>5\,R_\oplus$) in the catalog is a fraction of $\sim6\times10^{-4}$, which is modestly smaller than the currently inferred fraction of decay systems due to the faintness of Roman stars. As shown in Fig. \ref{fig:hist}, decay is highly sensitive to the value of $Q_{*}^{'}$, as the catalog yield goes from $0$ to $2$ to $\sim30$ when $Q_{*}^{'}$ goes from $10^7$ to $10^6$ to $10^5$.
\par The actual number of bulge fields in the survey will be $5$. Since the catalog only simulates the yield for one field, we expect the total yield to be $\sim5\times$ larger, our baseline estimate is that there will be $\sim5-10$ instances of orbital decay detectable across the entire survey.
\subsection{Alternative Parameterization}
\par
\begin{figure}[!t]
\centering
\includegraphics[width=0.47\textwidth]{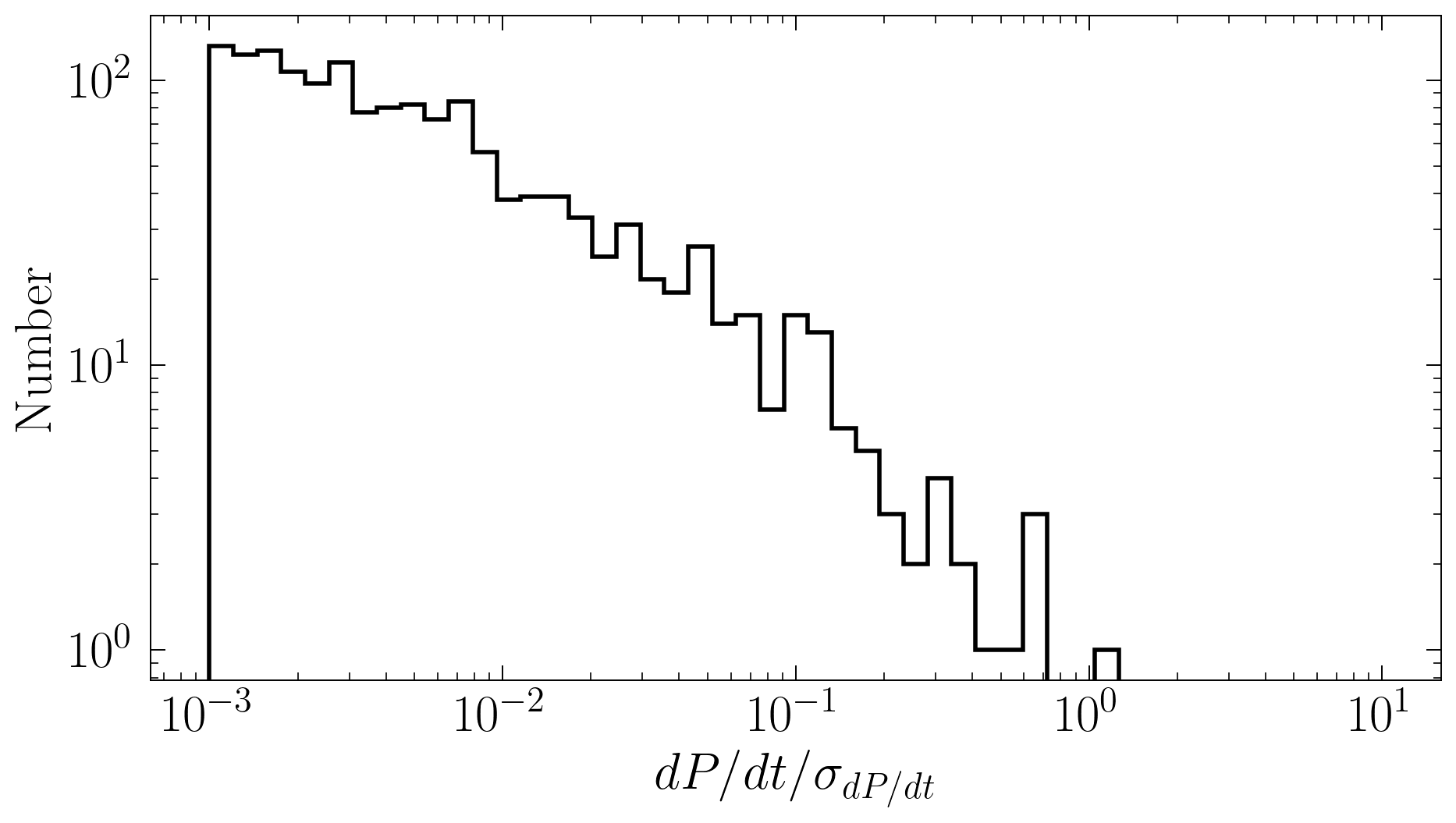}
\caption{The histogram of orbital decay SNRs $>10^{-3}$ using the $Q_{*}^{'}$ parameterization from \cite{millholland_empirical_2025}.}
\label{fig:histq}
\end{figure}

 The exact form of $Q_{*}^{'}$ is not well-constrained. \cite{millholland_empirical_2025} argued using the known population of short period planets that 
\begin{equation}
    Q_{*}^{'} = Q_{0}(P/2\,\textnormal{days})^{\alpha}, 
\end{equation}
with $-4.33	\lesssim\alpha\lesssim-2$ and $10^{5.5}\lesssim Q_0 \lesssim 10^7$, suggesting less efficient tidal dissipation at short periods, consistent with \cite{zanazzi_damping_2024}. We adopted $\alpha=-3$ and $Q_0 = 10^6$ and then re-estimate the SNR distribution, as shown in Fig. \ref{fig:histq}.
This form of $Q_{*}^{'}$ greatly reduces the decay rates for short periods to leave no detectable systems.

\subsection{Caveats}
\par Planet occurrence rates and properties in the Roman field might differ greatly from what we observe in the Solar neighborhood. Three of the four high-significance systems shown in Table \ref{tab:systems} are among the $\sim200$ catalog systems with log($g$)$\,< 3.8$, suggesting the yield of orbital decay is especially sensitive to how many transiting planets Roman observes orbiting subgiant stars. Roman will set lower limits on $Q_{*}^{'}$ for an unprecedentedly large sample of transiting planet hosts, improving empirical constraints on $Q_{*}^{'}$. 
\par For a variety of factors, our yield estimate is likely overly conservative. First, the \cite{wilson_transiting_2023} catalog used occurrence rates with only upper limits for much of the parameter space where $P \lesssim 1$ day. We expect that the catalog underestimates the number of planets that will be detected in the regime of interest for orbital decay. Additionally, the catalog does not simulate planets around giants, which Roman is expected to detect.
\par Orbital decay is not the only phenomenon that causes deviations from a linear ephemeris model. Precession, dynamical perturbations from other bodies, and acceleration of the system along our line of sight can lead to a non-zero period derivative. It often requires some detective work to decipher the physical origin of an inferred period derivative \citep[e.g.,][]{bouma_wasp-4_2020, ivshina_tess_2022}. Radial velocity measurements will be difficult to acquire for Roman stars so excluding line-of-sight acceleration directly is generally not feasible. The number of detectable period derivatives caused by line-of-sight acceleration depends on the number of widely-orbiting companions to planet-hosting stars and on their eccentricity distribution. Line-of-sight acceleration should produce systems whose periods are inferred to be increasing as well as decreasing so observing a substantial excess of inferred negative period derivatives would help establish orbital decay at a population-level. One can exclude apsidal precession by timing secondary eclipses \citep{yee_orbit_2020} or by inferring precession would require unphysically large planet Love numbers \citep{vissapragada_possible_2022}. Apsidal precession causes a sinusoidal, rather than quadratic, timing deviations so long-term transit time modeling might rule out precession as well. It might be difficult to establish orbital decay as the most likely cause of an inferred period derivative for any individual system.

\section{Conclusions} \label{sec:conc}
\par We expect that Roman will be a powerful laboratory for probing orbital decay, a key mechanism for sculpting planetary demographics in the small-period regime. We expect $5-10$ examples of orbital decay to be detectable over the entire duration of the GBTDS. If these detections do come to fruition, they would expand the number of firm orbital decay detections by a factor of 2 or more.
\par Developing a pipeline for transit time fitting and deviations from a linear ephemeris will be crucial to actually identifying feasible decay candidates. Robust vetting of planet candidates and decay candidates will be essential to eliminate sources of false positives. Regardless of the actual yield, Roman transiting planets will provide a large, homogeneous sample with which to constrain the stellar tidal dissipation factor $Q_{*}^{'}$ of planet host stars.
\section{Acknowledgments}
We would like to thank our anonymous referee, whose suggestions substantially improved the quality of this paper. K.C. would like to thank Luke Bouma, Sarah Millholland, and Christopher Kochanek for useful comments and feedback. K.C., B.S.G., and R.F.W. were supported by NASA Grant 80NSSC24K0917. B.S.G. was additionally supported by the Thomas Jefferson Chair Endowment for Discovery and Space Exploration, and K.C. was additionally supported by the Ohio State Distinguished University Fellowship. The material is based upon work supported by NASA under award number 80GSFC24M0006.

%

\vspace{5mm}


\software{astropy \citep{collaboration_astropy_2022}, numpy \citep{harris_array_2020}, pandas \citep{mckinney-proc-scipy-2010}, matplotlib \citep{hunter_matplotlib_2007} 
          }



\bibliography{orbdec}{}
\bibliographystyle{aasjournal}



\end{document}